%% file: TACAS19.tex
\documentclass[runningheads,a4paper]{llncs}
\usepackage{graphicx}
\usepackage{amsmath}
\usepackage{amssymb}
\usepackage{xspace}

\usepackage{array}

\usepackage[linesnumbered,ruled]{algorithm2e}

\usepackage{color}
\usepackage{xcolor}

\newif\ifArx
\Arxtrue  
\newcommand{\props}{P}
\newcommand{\model}{\mathcal{M}}

\newcommand{\closure}{\mathcal{C}}

\newcommand{\dist}[1]{\mathcal{D}^{#1}} 

\newcommand{\reals}{\mathbb{R}}
\newcommand{\nats}{\mathbb{N}}

\newcommand{\arel}{\mathcal{R}}

\newcommand{\topochecker}{{\ttfamily{topochecker}}\xspace}
\newcommand{\SLCS}{SLCS\xspace}
\newcommand{\SLCSMI}{ImgQL\xspace}
\newcommand{\ImgQL}{{\ttfamily{ImgQL}}\xspace}
\newcommand{\imgqool}{{\ttfamily{VoxLogicA}}\xspace}

\newcommand{\lnear}{\mathcal{N}}
\newcommand{\lsurr}{\mathcal{S}}
\newcommand{\lflood}{\mathcal{F}}
\newcommand{\lreach}{\rho}
\newcommand{\slreach}[2]{\lreach \;#1[#2]}

\ifArx
\SetAlFnt{\small}
\SetAlCapFnt{\small}
\SetAlCapNameFnt{\small}
\else
\SetAlFnt{\tiny}
\SetAlCapFnt{\scriptsize}
\SetAlCapNameFnt{\scriptsize}
\fi
\SetAlCapHSkip{0pt}
\IncMargin{-\parindent}

\newcommand{\imgLet}{{\ttfamily{let}}\xspace}
\newcommand{\imgOr}{\xspace{\ttfamily{|}}\xspace}
\newcommand{\imgAnd}{\xspace{\ttfamily{\&}}\xspace}
\newcommand{\imgImport}{{\ttfamily{import}}\xspace}
\newcommand{\imgNot}{\xspace{\ttfamily{!}}}
\newcommand{\imgLoad}{{\ttfamily{load}}\xspace}
\newcommand{\imgSave}{{\ttfamily{save}}\xspace}

\newcommand{\msep}{\,\mid\,}
\newcommand{\form}{{\bf F}}

\colorlet{VINC}{orange}
\colorlet{GINA}{blue}

\newcolumntype{L}{>{$}l<{$}}

\usepackage{subfig}
\usepackage{tikz}
\usepackage{latexsym}
\definecolor{darkgreen}{RGB}{30,120,30}

\newcommand{\RED}[1]{\textcolor{red}{#1}}
\newcommand{\BROWN}[1]{\textcolor{brown}{#1}}

\newcommand{\GREEN}[1]{\textcolor{darkgreen}{#1}}
\newcounter{dgnot}
\newenvironment{dgnot}[1][]{\refstepcounter{dgnot}\par\medskip
   \noindent \textbf{\RED{NfD~\thedgnot.}  #1} \rmfamily}{\medskip}
\newcommand{\nfd}[1]{\begin{dgnot} \RED{#1} \end{dgnot}}

\newcommand{\ed}{\hfill$\bullet$}
\newcommand{\ep}{\hfill$\diamond$}
\newcommand{\SET}[1]{\{#1\}}
\newcommand{\ZET}[2]{\SET{#1 | #2}}

\newcommand{\attrib}{A}
\newcommand{\aeval}{{\cal A}}
\newcommand{\peval}{{\cal V}}
\newcommand{\weval}{{\cal W}}
\newcommand{\lneg}{\neg}

\newcommand{\lssim}[7]{\triangle\!\!\!{\scriptstyle \triangle}_{#1}
{\tiny 
\left[
\begin{array}{ccc}
#2 & #3 & #4\\
#5 & #6 & #7
\end{array}
\right]
}}

\newcommand{\mkhis}{{\cal H}}
\newcommand{\mean}[1]{\overline{#1}}
\newcommand{\cc}{\mathbf{r}}

\renewcommand{\form}{\Phi}
\renewcommand{\arel}{R}

\usepackage{doi}
\usepackage{url}
\newcommand\code[1]{\texttt{#1}}

\newcommand{\MAGENTA}[1]{\textcolor{magenta}{#1}}

\newcounter{mknot}
\newenvironment{mknot}[1][]{\refstepcounter{mknot}\par\medskip
   \noindent \textbf{\MAGENTA{NfM~\themknot.}  #1} \rmfamily}{\medskip}
\newcommand{\nfm}[1]{\begin{mknot} \MAGENTA{#1} \end{mknot}}

\newcounter{vcnot}
\newenvironment{vcnot}[1][]{\refstepcounter{vcnot}\par\medskip
   \noindent \textbf{\BROWN{NfV~\thevcnot.}  #1} \rmfamily}{\medskip}
\newcommand{\nfv}[1]{\begin{vcnot} \BROWN{#1} \end{vcnot}}

\newcounter{gbnot}
\newenvironment{gbnot}[1][]{\refstepcounter{gbnot}\par\medskip
   \noindent \textbf{\GREEN{NfG~\thegbnot.}  #1} \rmfamily}{\medskip}

\providecommand{\url}[1]{{#1}}

\def\deriv{\noindent\hspace*{.20in}\vspace{0.1in}}

\def\contnl{\\$ $\mbox{\hspace{0.2in}}}

\def\hint#1#2{\newline#1\hspace*{.25in}\{$#2$\}\vspace{0.1in}\newline\hspace*{.20in}\vspace{0.1in}}

\newcommand{\brats}{BraTS 2017\xspace}

\newcommand{\old}[1]{}

\begin{document}

\mainmatter


\ifArx
\title{\imgqool: a Spatial Model Checker for Declarative Image Analysis (Extended Version)}
\else
\title{\imgqool: a Spatial Model Checker for Declarative Image Analysis}
\fi
\titlerunning{\imgqool}

\author{Gina Belmonte\inst{1}, Vincenzo Ciancia\inst{2} \and Diego~Latella\inst{2} \and Mieke~Massink\inst{2}}
\institute{Azienda Ospedaliera Universitaria Senese, Siena, Italy \and
Consiglio Nazionale delle Ricerche - Istituto di Scienza e Tecnologie dell'Informazione \lq A.~Faedo\rq, CNR, Pisa, Italy}

\authorrunning{Ciancia et al.}

\maketitle
\newcommand{\computer}{a desktop computer equipped with a 7th generation Intel Core I7 processor and 16GB of RAM\xspace}
\renewcommand{\nfm}[1]{}
\renewcommand{\nfv}[1]{}
\renewcommand{\nfd}[1]{}

\ifArx 
\nfm{TBD:
\begin{itemize}
\item Finalise Section "Efficiency comparison with topochecker"
\item Finalise case study results (Table!) [Done, 12/11/2018]
\item Finalise long arXiv version of the paper. See notes in long version.
\item Insert references to arXiv version, manual and STTT-arXiv paper, other refs?
\end{itemize}}
\nfm{\RED{IMPORTANT} Note that TACAS 2019 does \RED{NOT} allow annexes to be inserted (beyond 15 pages limit). Annexes can be in arXiv extended versions of the paper, which can be referred to in the submission. TBD!!}
\nfm{Version 17 of Nov. 09, 2018. Split TACAS (15pp) and ArXiv versions by if-then-else toggle.}
\nfm{Version 15 of Nov. 08, 2018. Revised mayreach operator and Specification.}
\nfm{Version 14 of Nov. 07, 2018. Revised all, but remaining issues and ca. 4 pages too long (besides annex and refs).}
\nfm{Version 13 of Nov. 06, 2018. Revised Case study section and abstract.}
\nfm{Version 10 of Oct. 30, 2018. Added some examples and start of results section.}
\nfv{Version 9 of Oct. 28, 2018. Added the Tool Section and some proposed modifications.}
\fi 

\begin{abstract}
Spatial and spatio-temporal model checking techniques
have a wide range of application domains, among which large scale distributed systems and signal and image analysis. We explore a new domain, namely (semi-)automatic contouring in Medical Imaging, introducing the tool \imgqool\ which merges the state-of-the-art library of computational imaging algorithms \code{ITK} with the unique combination of declarative specification and optimised execution provided by spatial logic model checking. 
The result is a \emph{rapid}, logic based analysis development methodology.
The analysis of an existing benchmark of medical images for segmentation of brain tumours shows that simple \imgqool\ analysis can reach state-of-the-art accuracy, competing with best-in-class
 algorithms, with the advantage of \emph{explainability} and \emph{replicability}. 
Furthermore, due to a two-orders-of-magnitude speedup compared to the existing \emph{general-purpose} spatio-temporal model checker \topochecker, \imgqool\ enables {\em interactive} development of analysis of 3D medical images, which can greatly facilitate the work of professionals in this domain.

\end{abstract}

\begin{keywords}
Spatial logics; Closure spaces; Model
checking; Medical Imaging;
\ifArx 
Segmentation; Magnetic
Resonance Imaging; Distance Transform; Statistical
Texture Analysis
\fi
\end{keywords}


\input{Introduction}

\input{SpatialLogicFramework}
\input{Tool}

\input{Conclusions}

\ifArx 
\appendix

\section{Proof of Proposition~\ref{prop:SurrFromReachSimple}}

\begin{proof}
We prove that 
$$
\model, x \not\models \form_1 \, \lsurr \, \form_2
\mbox{ if and only if }
\model, x \not\models \form_1 \, \land\lneg 
(\slreach{\lneg(\form_1 \lor \, \form_2)}{\lneg \form_2})
$$
where by
$\model, x \not\models \form$
we mean that
$\model, x \models \form$  does not hold:\\

\noindent
$
\deriv
\model, x \not\models \form_1 \, \land\lneg 
(\slreach{\lneg(\form_1 \lor \, \form_2)}{\lneg \form_2})
\hint{\Leftrightarrow}{Defs. of $\not\models$, $\land$; Logic}
\model, x \not\models \form_1 \mbox{ or } \contnl
\model, x \not\models \lneg 
(\slreach{\lneg(\form_1 \lor \, \form_2)}{ \lneg \form_2})\\
\hint{\Leftrightarrow}{Defs. of $\not\models$, $\lneg$}
\model, x \not\models \form_1 \mbox{ or }  \contnl
\model, x \models
\slreach{\lneg(\form_1 \lor \, \form_2)}{ \lneg \form_2}\\
\hint{\Leftrightarrow}{Def. of $\lreach$}
\model, x \not\models \form_1\mbox{ or }  \contnl
\begin{array}{l l l}
\mbox{exist path} &\pi  \mbox{ and index }  \ell  \mbox{ s.t. :}   &\\
& \pi(0)=x \,\mbox{ and }  \\
& \model, \pi(\ell) \models \lneg (\form_1 \lor \, \form_2)\,\mbox{ and } \\
& \mbox{for all } j : 0 < j < \ell \mbox{ implies }\model,\pi(j) \models  \lneg \form_2
\end{array}\\
\hint{\Leftrightarrow}{Defs. of $\lneg$, $\lor$, $\not\models$;  Logic}
\model, x \not\models \form_1 \mbox{ or } \contnl
\begin{array}{l l l}
\mbox{exist path} & \pi  \mbox{ and index } \ell  \mbox{ s.t. :} \\
  & \pi(0)=x \,\mbox{ and } \\
& \model, \pi(\ell) \models \lneg\form_1 \mbox{ and } \\
& \model, \pi(\ell) \models \lneg\form_2 \mbox{ and } \\
& \mbox{for all } j: 0 < j < \ell \mbox{ implies }\model,\pi(j) \models  \lneg\form_2
\end{array}\\
\hint{\Leftrightarrow}{Logic}
\model, x \not\models \form_1 \mbox{ or } \contnl
\begin{array}{l l l}
\mbox{exist path} & \pi  \mbox{ and index } \ell  \mbox{ s.t. :} \\
 & \pi(0)=x \,\mbox{ and } \\
& \model, \pi(\ell) \models \lneg\form_1 \mbox{ and } \\
& \mbox{for all } j : 0< j \leq \ell \mbox{ implies } \model,\pi(j) \models  \lneg \form_2
\end{array}\\
\hint{\Leftrightarrow}{Defs. of $\not\models$, $\lsurr$}
\model, x \not\models \form_1 \, \lsurr \, \form_2
$
\ep
\end{proof}

\nfv{OCCHIO:The following definitions are instrumental for the proof of correctness of the implementation of the
model-checking procedure for the flood operator $\lflood$. Note that the flood operator is not the same as the $\rho$-operator. Moreover, it may be that below we intended the "reverse" flood operator. \RED{The connection to the connected components-based algorithm still has to be proven/developed.}
Non ci e' chiaro che farcene 
\RED{(basta un commento nel codice? \ldots )}}

\bibliographystyle{splncs03}
\bibliography{TACAS19.bib}

\end{document}

%% file: Introduction.tex

\section{Introduction and Related Work}
\label{sec:Introduction}

\emph{Spatial  and Spatio-temporal model checking} have gained an increasing interest in recent years in various domains of application ranging from Collective Adaptive Systems~\cite{CLMP15,CGLLM14,CLMPV16} and networked systems~\cite{SPATEL}, to signals~\cite{NBCLM15} and images~\cite{Gr+09,CLLM16}. Research in this field has its origin in the \emph{topological} approach to spatial logics, dating back to the work of Alfred Tarski, who first recognised the possibility of reasoning on physical, continuous, space using topology as a mathematical framework for the interpretation of modal logic (see~\cite{Ch5HBSL} for a thorough introduction). More recently these early theoretical foundations have been extended to encompass reasoning about {\em discrete} spatial structures, such as graphs and images, extending the theoretical framework of topology to {\em (quasi discrete) closure spaces} (see for instance~\cite{Gal99,Gal03,HBSL}). 
That framework has subsequently been taken further in recent work by Ciancia et al.~\cite{CLLM14,CLLM16,CLLM16bertinoro} 
resulting in the definition of the \emph{Spatial Logic for Closure Spaces} (\SLCS), temporal extensions (see \cite{CGLLM15,TKG17,NBCLM15}), and related model checking algorithms and tools.

The main idea of spatial (and spatio-temporal) model checking is to use specifications written in a suitable logical language to describe spatial properties and to automatically identify patterns and structures of interest in a variety of domains (see e.g., \cite{Bartocci2016,CLMPV16,Ciancia2018}). In this paper we focus on one such domain, namely medical imaging for radiotherapy, and brain tumour segmentation in particular, which is a important and currently very active research domain of its own. One of the technical challenges of the development of automated (brain) tumour segmentation is that lesion areas are only defined through differences in the intensity (luminosity) in the (black \& white) images that are {\em relative} to the intensity of the surrounding normal tissue. A further complication is that even (laborious and time consuming) manual segmentation by experts shows significant variations when intensity gradients between adjacent tissue structures are smooth or partially obscured~\cite{Menze2015}. Moreover, there is a considerable variation across images from different patients and images obtained with different Magnetic Resonance Images (MRI) scanners. Several automatic and semi-automatic methods have been proposed in this very active research area (see for example~\cite{lemieux1999fast,Zhu2012,Despotovi2015,Simi2015,Dupont2016,Fyllingen2016}).

In this paper, continuing the research line of \cite{BCLM16,Ba+19,Be+17}, we present the free and open source tool \imgqool (acronym for \emph{Voxel-based Logical Analyser})\footnote{\imgqool: \url{https://github.com/vincenzoml/VoxLogicA}}, catering  for a novel approach to image segmentation, namely a \emph{rapid-development}, declarative, logic-based method, supported by {\em spatial model checking}, tailored to identify `regions of interest' in medical images.
This approach is particularly suitable to reason at the ``macro-level'', by exploiting the {\em relative} spatial relations between tissues or organs at risk. 
\imgqool\ is similar, in the accepted logical language, and functionality, to the spatio-temporal model checker \topochecker\footnote{Topochecker: \emph{a topological model checker}, see \url{http://topochecker.isti.cnr.it}, \url{https://github.com/vincenzoml/topochecker}}, but \old{it has been rewritten from scratch, and} specifically designed for the analysis of (possibly multi-dimensional, e.g. 3D) \emph{digital images} as a specialised image analysis tool. It is tailored to usability and efficiency by employing state-of-the-art algorithms and open source libraries, borrowed from computational image processing, in combination with efficient spatial model checking algorithms.

We show the application of \imgqool\ on \brats~\cite{Menze2015,Bak+17}, a publicly available set of benchmark MRI images for brain tumour segmentation, with an associated yearly challenge. For each image in the benchmark a manual segmentation of the tumour by domain experts is available, enabling rigorous and objective qualitative comparisons via established similarity indexes. 
%
We show that our simple, high-level logical specifications for glioblastoma segmentation are directly competitive with the state-of-the-art techniques submitted to the \brats challenge, some of which based on machine learning. Our approach to segmentation has the unique advantage of \emph{explainability} and {\em replicability}. Logically specified procedures can be understood by humans and improved to encompass new observations by domain experts. The fact that our segmentation results are of very high quality indicates, in our opinion, that in the medical domain, expert knowledge -- that is central in our approach -- plays a fundamental role. However, machine learning can still be used, for instance, to fine-tune numeric parameters for the analysis, and will constitute exciting future work.

\ifArx 
The outline of the paper is as follows. Section~\ref{sec:SpatialLogicFramework} briefly recalls the spatial logic framework on which \imgqool\ is based. Section~\ref{sec:Tool} describes the architecture of the \imgqool\ and an example of its use, whereas in Section~\ref{sec:case-study} the results of the application of \imgqool\ on a publicly available set of benchmark MRI images are presented. Finally, in Sect.~\ref{sec:Conclusions} we conclude and provide an outline for future research.
\fi 

%% file: SpatialLogicFramework.tex
\section{The Spatial Logic Framework}
\label{sec:SpatialLogicFramework}
In this section,  we  briefly recall the logical language \SLCSMI (\emph{Image Query Language}) 
proposed in~\cite{Ba+19}, which is based on the \emph{Spatial Logic for Closure Spaces} \SLCS \cite{CLLM14,CLLM16} and which forms the {\em kernel} of the framework
we propose in the present paper. We then show two examples of operators specifically designed for digital image analysis, which we express as additional logic operators. In Section~\ref{sec:case-study} we will show how the
resulting logic can be used for actual analysis via spatial model checking. 

\subsection{Foundations: Spatial Logics for Closure Spaces}
\label{sec:Kernel}
The logic for closure spaces we use in the present paper is closely related to
\SLCS~\cite{CLLM14,CLLM16} and, in particular, to the \SLCS extension with 
distance-based  operators presented in~\cite{Ba+19}. As in~\cite{Ba+19}, the resulting
logic constitutes the {\em kernel} of a solid logical framework  
for reasoning about texture features of digital 
images, when interpreted as closure spaces. 

In the context of our work, a {\em digital image} is not only a 2-dimensional grid of \emph{pixels}, but, more generally, a multi-dimensional (very often, 3-dimensional) grid of hyper-rectangular elements that are called \emph{voxels} (``volumetric picture elements''). When voxels are not \emph{hypercubes}, images are said to be \emph{anisotropic}; this is usually the case in medical imaging. Furthermore, a digital image may contain information about its ``real world'' spatial dimensions, position (origin) and rotation, permitting one to compute the real-world coordinates of the centre and edges of each voxel. In medical imaging, such information is typically encapsulated into data by machines such as MRI scanners. In the remainder of the paper, we make no dimensionality assumptions, and we therefore refer to picture elements as voxels.
We recall the main definitions below referring the reader to the afore mentioned papers for details, examples and discussion.

\begin{definition}\label{def:ClosureSpaces}
A {\em closure space} is a pair $(X,\closure)$ where $X$ is a non-empty set (of {\em points})
and $\closure: 2^X \to 2^X$ is a function satisfying the following axioms:
$\closure(\emptyset)=\emptyset$;
$Y \subseteq \closure(Y)$ for all  $Y \subseteq X$;
$\closure(Y_1 \cup Y_2) = \closure(Y_1) \cup \closure(Y_2)$ for all $Y_1,Y_2\subseteq X$. \ed
\end{definition}

Given any relation $\arel \subseteq X \times X$, function $\closure_{\arel}:2^X \to 2^X$ 
with $\closure_{\arel}(Y) \triangleq Y\cup \ZET{x}{\exists y \in Y. y \, \arel \,x}$ satisfies the axioms of Definition~\ref{def:ClosureSpaces} thus making 
$(X,\closure_{\arel})$ a closure space. Whenever a closure space is generated by a relation as above, it is called a {\em quasi-discrete} closure space.
A quasi-discrete closure space $(X,\closure_{\arel})$, can be used as the
basis for a mathematical model of a digital image.
$X$ represents the finite set
of {\em voxels} and $\arel$ is the reflexive and symmetric  {\em adjacency} relation between voxels~\cite{Gal14}. 
\ifArx 
We note in passing that several different adjacency relations can be used.
For instance, in the {\em orthogonal}  adjacency relation only voxels which share an edge count as adjacent, so that, in the 2D case, each pixel is adjacent  to (itself and) four other pixels; on the other hand, in the {\em orthodiagonal}  adjacency relation voxels are adjacent as long as they share at least either an edge or a corner,
so that, again in the 2D case, each pixel is adjacent  to (itself and) eight other pixels. 
\fi 
A closure space $(X,\closure)$ can be enriched with a notion of {\em distance}, i.e. a function $d: X \times X \to \reals_{\geq 0} \cup \SET{\infty}$ such that $d(x,y)=0$ iff $x=y$, leading to the {\em distance closure space} $((X,\closure),d)$.\footnote{We recall that  for 
$\emptyset\not=Y\subseteq X$, $d(x,Y) \triangleq  \inf \ZET{d(x,y)}{y \in Y}$, with $d(x,\emptyset) =\infty$.
In addition, as the definition of $d$ might require the elements of $\arel$ to be weighted, 
quasi-discrete distance closure spaces
may be enriched with a $\arel$-weighting function $\weval:\arel \to \reals$
assigning the weight $\weval(x,y)$ to each $(x,y)\in \arel$. In the sequel we will keep 
$\weval$ implicit, whenever possible and for the sake of simplicity.} 

It is sometimes convenient to equip the points of a closure space with {\em attributes};
for instance, in the case of images, such attributes could be the color or intensity of voxels.
We assume sets $A$ and $V$ of attribute {\em  names} and {\em values}, and 
an {\em attribute valuation} function $\aeval$ such that $\aeval(x,a) \in V$ is the value of attribute $a$ of point $x$.
Attributes can be used in {\em assertions} $\alpha$, i.e. boolean expressions, with standard syntax and semantics. Consequently, we abstract from related details here and assume function $\aeval$ extended in the obvious way; for instance, $\aeval(x,a \leq c) = \aeval(x,a) \leq c$, for appropriate constant $c$.

A (quasi-discrete) {\em path} $\pi$ in  $(X,\closure_{\arel})$ is a function
$\pi : \nats \to X$, 
such that for all $Y \subseteq \nats$, $\pi(\closure_{Succ}(Y)) \subseteq \closure_{\arel}(\pi(Y))$,
where  $(\nats, \closure_{Succ})$ is the closure space of natural numbers
with the {\em successor} relation: $(n,m) \in Succ \Leftrightarrow m=n+1$. Informally: the ordering in the path imposed by $\nats$ is compatible with relation $\arel$, i.e. $\pi(i) \, \arel\, \pi(i+1)$.
%
%
%
%
%
For given set $\props$ of {\em atomic predicates} $p$, and interval of $\reals$ $I$, the syntax of the logic 
we use in this paper is given below:
\begin{equation}\label{def:syntax}
\form  ::=   p  \msep  \lneg \, \form  \msep  \form_1 \, \land \, \form_2  \msep  \lnear \form \msep \slreach{\form_1}{\form_2} \msep \dist{I} \form 
\end{equation}
Informally,  it is assumed that space is modelled by the set of points of a distance closure model; each atomic predicate  $p \in \props$ models a specific {\em feature} of {\em points} and is thus associated with the points
that have this feature\footnote{In particular, a predicate $p$ can be a {\em defined} one, by means 
of a definition as $p:=\alpha$, meaning that the feature of interest is characterized by the (boolean) value of $\alpha$.}. A point $x$ satisfies  $\lnear\, \form$ if a point satisfying $\form$
can be reached from $x$ in at most one (closure) step, i.e. if $x$ is {\em near} (or {\em close}) to
a point satisfying  $\form$; $x$ satisfies $\slreach{\form_1}{\form_2}$
if $x$ {\em may reach} a point satisfying $\form_1$ via a path passing only by points 
satisfying $\form_2$; it satisfies $\dist{I} \, \form$ if its distance from the set of points satisfying 
$\form$ falls in interval $I$. 
Finally, the logic includes logical negation ($\neg$) and 
conjunction ($\land$).
In the following we formalise the semantics of the logic, noting that all definitions are independent from the nature of the paths (e.g. being quasi-discrete or not) or of the closure space.

\begin{definition}\label{def:model}
A {\em distance closure model}  $\model$ is a tuple $\model=(((X,\closure),d), \aeval, \peval)$, where $((X,\closure),d)$ is a distance closure space, $\aeval: X \times \attrib \to V$ an attribute valuation, and $\peval: \props \to 2^X$ is a valuation  assigning to
each atomic predicate the set of points where it holds.
\ed
\end{definition}

\begin{definition}\label{def:satisfaction}
{\em Satisfaction} $\model, x \models \form$  of a formula $\form$ at point $x \in X$ in model
$\model = (((X,\closure),d), \aeval, \peval)$ is defined by induction on the structure of formulas:
\[
\begin{array}{r c l c l c l L}
\model,x & \models  & p \in P & \Leftrightarrow & x  \in \peval(p)\\
\model,x & \models  & \lneg \,\form & \Leftrightarrow & \model,x  \models \form \mbox{ does not hold}\\
\model,x & \models  & \form_1\, \land \,\form_2 & \Leftrightarrow &
\model,x  \models \form_1 \mbox{ and } \model,x  \models \form_2\\
\model,x & \models  & \lnear \, \form & \Leftrightarrow & x \in \closure(\ZET{y}{\model,y \models \form})\\
\model,x & \models  & \slreach{\form_1}{\form_2} & \Leftrightarrow & \mbox{there exist path } \pi 
\mbox{ and index } \ell 
\mbox{ such that:}\\
 & &  & & 
\mbox{\hspace{0.1in}}\pi(0)=x \mbox{ and } 
\mbox{\hspace{0.1in}}\model, \pi(\ell) \models  \form_1 \mbox{ and }\\
 & &  & & 
\mbox{\hspace{0.1in}}\mbox{for all indexes } j: 0 < j  < \ell \mbox{ implies } \model, \pi(j) \models \form_2\\
\model, x & \models & \dist{I} \, \form & \Leftrightarrow &
d(x, \ZET{y}{\model, y \models \form}) \in I
\end{array}
\]
where, whenever $p:=\alpha$ is a definition for $p$, we assume $x\in \peval(p)$ if and only if $\aeval(x,\alpha)$ yields the truth-value $true$. 
\ed
\end{definition}

In the logic proposed in~\cite{CLLM14,CLLM16}, the ``may reach'' operator is not present, and the {\em surrounded}  
operator $\lsurr$ has been defined as  basic operator as follows:
$x$ satisfies $\form_1 \, \lsurr \, \form_2$ if and only if $x$ belongs to an area
of points satisfying $\form_1$ and one cannot ``escape'' from such an area 
without hitting a point satisfying $\form_2$.
Several types of \emph{reachability} predicates can be derived from $\lsurr$. However, reachability is in turn a widespread primitive, implemented in various forms (e.g., \emph{flooding}, \emph{connected components}) in programming libraries. Thus, in this work we prefer to use reachability as a basic predicate of the logic\footnote{Similar considerations are also present in \cite{BBLN17} -- extending the combination of Signal Temporal Logic with spatial operators \emph{a la \SLCS} first presented in \cite{NBCLM15}.
}. In the sequel we show that $\lsurr$ can be derived from the operators defined above,
employing a definition patterned after the model-checking algorithm of \cite{CLLM14}.
This change simplifies the definition of several derived connectives, including that of \emph{touch} (see below), and resulted in notably faster execution times for analyses using such derived connectives.
We first recall the formal definition of $\lsurr$:
\[
\begin{array}{r c l c l c l L}
\model,x & \models  & \form_1 \, \lsurr \, \form_2 & \Leftrightarrow & \model,x \models  \form_1 \, \mbox{\em and}\\
 & &  & & \mbox{\em for all paths } \pi 
 \mbox{\em and indexes } \ell \mbox{ \em the following holds:}\\
 & &  & & 
\mbox{\hspace{0.1in}}\pi(0)=x \, \mbox{\em and } \,  \model, \pi(\ell) \models \lneg \form_1\\
 & &  & & 
\mbox{\hspace{0.1in}}\mbox{\em implies }\\
 & &  & &
\mbox{\hspace{0.1in}}\mbox{\em there exists index } j \mbox{\em such that:}\\
& &  & &
\mbox{\hspace{0.3in}}0 < j \leq \ell \, \mbox{\em and } \,  \model, \pi(j) \models \form_2
\end{array}
\]
\begin{proposition}\label{prop:SurrFromReachSimple}
For all 
closure models  $\model=((X,\closure), \aeval, \peval)$ 
and all formulas $\form_1$, $\form_2$ the following holds:
$
\form_1 \, \lsurr \, \form_2  \equiv 
\form_1 \, \land\lneg 
(\slreach{\lneg(\form_1 \lor \, \form_2)}{\lneg \form_2})
$
\ep
\end{proposition}
\ifArx 
\nfd{Sebbene nella dimostrazione io avessi assunto il modello quasi-discreto a me pare che essa valga per qualunque modello per il quale abbiamo una nozione di path! Quindi ho tolto 
i riferimenti alla quasi-discrete \ldots}
\fi 

\begin{definition}\label{def:derived}
We define a number of derived operators that are of particular use in medical image analysis:
\[
\begin{array}{r c l c l c l L}
\model,x & \models & \mathit{touch } (\form_1 , \form_2) \Leftrightarrow \form_1 \land \slreach{\form_2}{\form_1}\\
\model,x & \models  & \mathit{grow } (\form_1 , \form_2)  \Leftrightarrow \form_1 \lor \mathit{touch}( \form_2 , \form_1)\\
\model,x & \models  & \mathit{flt }(r, \form_1) \Leftrightarrow \dist{\leq r} (\dist{\geq r} \lnot\form_1)
\end{array}
\]
\ed
\end{definition}

The formula $\mathit{touch } (\form_1 , \form_2)$ is satisfied by points that satisfy $\form_1$ and that are on a $\form_1$-path that can reach a point satisfying $\form_2$. The formula  $\mathit{grow } (\form_1 , \form_2)$ is satisfied by points that satisfy $\form_1$ and by points that satisfy $\form_2$ which are on a $\form_2$-path that can reach a point satisfying $\form_1$.  The formula $\mathit{flt }(r,\form_1)$ is satisfied by points that are at a distance of less than $r$ from a point that is at least at distance $r$ from points that do not satisfy $\form_1$. This operator in practice works as a filter where only contiguous areas of points satisfying $\form_1$ that have a minimal diameter of at least $2*r$ are preserved (or taken into consideration), while also smoothening those areas in case they have an irregular shape (e.g. having protrusions of less than the indicated distance).

\begin{figure}
\centering
\subfloat[][]
{
\includegraphics[height=1.5cm]{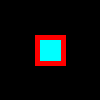}
\label{subfig:surround_test3}
}\hfill
\centering
\subfloat[][]
{
\includegraphics[height=1.5cm]{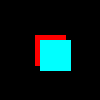}
\label{subfig:surround_test4}
}\hfill
\centering
\subfloat[][]
{
\includegraphics[height=1.5cm]{img/MMtest4.png}
\label{subfig:surround_test4b}
}\hfill
\centering
\subfloat[][]
{
\includegraphics[height=1.5cm]{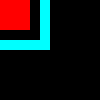}
\label{subfig:surround_test6}
}\\
\subfloat[][]
{
\includegraphics[height=1.5cm]{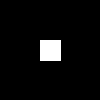}
\label{subfig:out_test3}
}\hfill
\centering
\subfloat[][]
{
\includegraphics[height=1.5cm]{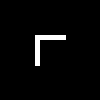}
\label{subfig:out_touchT4}
}\hfill
\centering
\subfloat[][] 
{
\includegraphics[height=1.5cm]{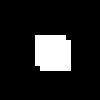}
\label{subfig:out_growT4}
}\hfill
\centering
\subfloat[][] 
{
\includegraphics[height=1.5cm]{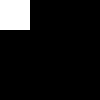}
\label{subfig:out_test6c}
}
\caption{Examples of operators surround, touch, grow and distance. Fig.~\ref{subfig:out_test3} $\mbox{blue } \lsurr \mbox{ red}$ of (a); Fig.~\ref{subfig:out_touchT4} $\mbox{touch}(\mbox{red},\mbox{blue})$ of (b); Fig.~\ref{subfig:out_growT4} $\mbox{grow}(\mbox{red},\mbox{blue})$ of (c); Fig.~\ref{subfig:out_test6c} $\mbox{red } \lsurr\; ( \dist{\le 11} \mbox{ blue})$ of (d). }
\label{fig:surround}
\end{figure}

We conclude this section with an  illustration of the derived spatial operators $\mathit{touch }$, $\mathit{grow }$ and the surrounded operator in combination with the distance operator. In the top row of
Fig.~\ref{fig:surround} some sample input images (bitmaps) are shown of 100 by 100 points that are red, blue or black. In the second row of Fig.~\ref{fig:surround} the results of  some spatial properties are shown, where points in the original image that satisfy the spatial formula $\phi$ are shown as white points. 
\ifArx 
Fig.~\ref{subfig:out_test3} shows the points that satisfy $\Phi=\mbox{blue } \lsurr \mbox{ red}$, i.e. blue points that are surrounded by red points in Fig.~\ref{subfig:surround_test3}. 
Fig.~\ref{subfig:out_touchT4} shows the points that satisfy $\Phi=\mbox{touch}(\mbox{red},\mbox{blue})$, i.e. red points that are on a path, consisting of only red points, that reaches a blue point in Fig.~\ref{subfig:surround_test4}. There are no points that satisfy $flt(5.0,\mbox{red})$ as the red area has a minimal diameter which is less than 5.0 (result not shown).
Fig.~\ref{subfig:out_growT4} shows the points that satisfy $\Phi=\mbox{grow}(\mbox{red},\mbox{blue})$ of those in Fig.~\ref{subfig:surround_test4b}.
Fig.~\ref{subfig:out_test6c} shows the results of a formula combining the surround operator with a distance operator where the white points are those corresponding to red points in Fig.~\ref{subfig:surround_test6} that are surrounded by points that are less than 11 points (in distance) away from points that are blue. In Fig.~\ref{subfig:surround_test6} the area with red points is 30 by 30 points, the black area separating the red and the blue areas is 10 points wide. 
\else 
The examples are self-explanatory. For more details the reader is referred to~\cite{TACASarxiv}.
\fi 

\subsection{Digital Image Analysis Specific Operators}
\label{sec:DIASpecificOperators}

In the sequel, we provide some details on a logical operator, first defined in \cite{Ba+19},
that we use in the context of Texture Analysis (see for example~\cite{Kassner2010,Lopes2011,Castellano2004,Davnall2012}) 
for defining a notion of {\em statistical similarity} between image regions. 
The statistical distribution of an area $Y$ of a black and white  image is
approximated by the \emph{histogram} of the grey levels of points (voxels) 
belonging to $Y$, limiting the representation
to those levels laying in a certain interval $[m,M]$, the latter being split into $k$ {\em bins}. In the case of images modelled as closure models, where each point may have several attributes, the histogram can be defined for different attributes. Given a closure model 
$\model = ((X,\closure), \aeval, \peval)$, define function 
$\mkhis : A \times 2^X \times \reals \times \reals \times \nats \to (\nats \to \nats)$
such that for all $m<M$, $k>0$ and $i\in \SET{1,\ldots,k}$,
$
\mkhis(a,Y,m,M,k)(i) =
\Big|\ZET{y\in Y}{(i-1) \cdot \Delta \leq \aeval(y,a) - m < i \cdot\Delta}\Big|
$
where $\Delta=\frac{M-m}{k}$. 
We call $\mkhis(a,Y,m,M,k)$  the {\em histogram} of $Y$ (for attribute $a$), 
with $k$ bins and $m$, $M$ min and max values respectively.  The {\em mean}
$\mean{h}$ of a histogram  $h$ with $k$ {\em bins} is the quantity
$\frac{1}{k}\sum_{i=1}^k h(i)$. The  {\em cross correlation} between two histograms 
$h_1, h_2$  with the same number $k$ of {\em bins} is defined as follows:
$$
\cc(h_1,h_2) = 
\frac
{\sum_{i=1}^k\left( h_1(i) - \mean{h_1} \right) \left(h_2(i) - \mean{h_2} \right)}
{
\sqrt{ 
\sum_{i=1}^k \left(h_1(i) - \mean{h_1} \right)^2
}
\sqrt{
\sum_{i=1}^k \left(h_2(i) - \mean{h_2} \right)^2
}
}
$$

The value of $\cc{}{}$ is \emph{normalised} so that
$-1\le \cc{}{}\le 1$; $\cc(h_1,h_2)=1$ indicates that $h_1$ and $h_2$
are \emph{perfectly correlated} (that is, $h_1 = ah_2+b$, with $a>0$); $\cc(h_1,h_2) =-1$
indicates \emph{perfect anti-correlation} (that is, $h_1=ah_2+b$, with $a<0$). On the other
hand, $\cc(h_1,h_2) = 0$ indicates no correlation. 

We embed {\em statistical similarity} $\lssim{\bowtie c}{m}{M}{k}{r}{a}{b}$ in
the logic by adding it to the grammar defined by (\ref{def:syntax}) and extending the definition of the satisfaction relation (Def.~\ref{def:satisfaction}) with the following equation,
with $m,M,k$ as above:
$$
\model, x \, \models \,
\lssim{\bowtie c}{m}{M}{k}{r}{a}{b} \form \Leftrightarrow
\cc(h_a,h_b) \bowtie c
$$
where
$h_a=\mkhis(a,S(x,r),m,M,k)$,
$h_b=\mkhis(b,\ZET{y}{\model,y \models \form},m,M,k)$,
$c$ is a constant in $[-1,1]$,
$\bowtie\, \in\SET{<, \leq, =, \geq, >}$  and
$S(x,r)=\ZET{y\in X}{d(x,y) \leq r}$ is the {\em sphere} of radius $r$ centred in $x$. Note that, differently from \topochecker that was used in \cite{Ba+19}, in \imgqool, for efficiency reasons, $r$ is actually the \emph{hypercube} with edge size $2*r$, which is discretely approximated by a hyperrectangle after computing the number of voxels corresponding to $r$ in each dimension.

So $\lssim{\bowtie c}{m}{M}{k}{r}{a}{b} \form$ compares the region of the
image constituted by the sphere (hypercube) of radius $r$ centred in $x$ against the 
region characterised by $\form$. The comparison is  based on the cross correlation
of the histograms of the chosen attributes of (the points of) the two regions, namely $a$ and $b$
and both histograms share the same range ($[m,M]$) and the same bins ($[1,k]$). 
In summary, the operator allows to check {\em to which extent} the {\em sphere (hypercube) around the point of interest} is {\em statistically similar} to a given region (specified by) $\form$.

Finally, we introduce the {\em percentiles} operator; $\mathit{percentiles}(\phi_1,\phi_2)$ takes a number-valued image $\phi_1$ and a boolean-valued \emph{mask} $\phi_2$ and returns an image in which each point is associated to the {\em percentile rank} of its intensity in $\phi_1$ with respect to the population of voxels that are true in $\phi_2$. 
\ifArx 
For example, a point $x$ with percentile rank 0.8 is a point with an intensity such that 80\% of the points in the image, that correspond to the mask, have an intensity below that of point $p$. This quantitative operator is essential to be able to use the same segmentation specification on images that may vary in intensity distribution (i.e. a similar phenomenon as variations in slightly under or over exposed photographs). It makes it possible to avoid the use of absolute values in constraints on the intensity of points, as will be illustrated in Sect.~\ref{sec:case-study}.   
\fi 
The percentile ranking of a point is defined as $pr(p)=\frac{c_l(p)+0.5*f_i(p)}{N}* 100\%$, where $c_l(p)$ is the count of all intensities that are below the intensity of $p$, $f_i(p)$ is the frequency of the intensity of $p$ and $N$ is the total number of points in the considered part of the image (namely that part corresponding to the mask).


%% file: Tool.tex

\section{The Tool \imgqool}\label{sec:Tool}

Functionality-wise, \imgqool\ specialises \topochecker to the case of spatial analysis of {\em multi-dimensional images}. 
It interprets a specification written in the \SLCSMI language, using a set of multi-dimensional images\footnote{Besides common bitmap formats, the model loader of \imgqool\ currently supports the NIfTI (Neuro-imaging Informatics Technology Initiative) format (https://nifti.nimh.nih.gov/, version 1 and 2).
3D MR-FLAIR images in this format very often have a slice size of 256 by 256 pixels, multiplied by 20 to 30 slices.}
as models of the spatial logic, and produces as output a set of multi-dimensional images representing the valuation of user-specified expressions. For logical operators, such images are Boolean-valued, that is, \emph{regions of interest} in medical imaging terminology, which may be loaded as \emph{overlays} in medical image viewers. Non-logical operators may generate number-valued images. \imgqool augments \SLCSMI with file loading and saving primitives, and a set of additional commodity operators, specifically aimed at image analysis, that is destined to grow along with future developments of the tool.
The main execution modality of \imgqool is \emph{batch execution}. A (currently experimental) \emph{graphical user interface} is under development. 
\ifArx
A planned future development is \emph{interactive execution}, in particular for semi-automated analysis, by letting a domain expert calibrate numeric parameters in real-time, while seeing the intermediate and final results.
\fi

Implementation-wise, the tool achieves a two-orders-of-magnitude speedup with respect to \topochecker. Such speedup has permitted the rapid development of a novel procedure for automatic segmentation of \emph{glioblastoma} that, besides being competitive with respect to the state-of-the-art in the field (see Section \ref{sec:case-study}), is also easily \emph{replicable} and \emph{explainable} to humans, and therefore amenable of improvement by the community of medical imaging practitioners.

\subsection{Functionality}

We provide an overview of the tool functionality, starting from the syntax of analysis session files. For space reasons, we largely omit details on parsing rules (e.g. the syntax of identifiers and infix operators, which is delegated to the tool documentation).
In the following, \code{f, x1,\ldots, xN, x} are identifiers, 
\code{"s"} is a string, and \code{e1, \ldots, eN, e}
are expressions (to be detailed later).  A \imgqool specification consists of a text file containing a sequence of (side-effecting) \textbf{commands} (see Specification~\ref{alg:seg} in Sect.~\ref{sec:case-study} as an example). Five commands are currently implemented:

{\small
\begin{itemize}
 \item 
 \code{let f(x1,...,xN) = e} is used for \emph{function declaration}, also in the form \code{let f = e}\emph{(constant declaration)}, and with special syntactic provisions to define \emph{infix} operators; declarations have no imperative side effects, and only affect name bindings, so that after execution of the command, name \code{f} is bound to a function or constant that evaluates to \code{e} -- with the appropriate substitutions of expressions (to be detailed later) for parameters;
 
 \item
 \code{load x = "s"}, loads an image from file \code{"s"} and binds it to \code{x} for subsequent usage; 
 
 \item
 \code{save "s" e}, stores the image resulting from evaluation of expression \code{e} to file \code{"s"};
 
 \item
 \code{print "s" e} prints to the log \ifArx (in batch mode, the console, with prints decorated by elapsed time and other debug information)\fi the string \code{s} followed by the numeric, or boolean, result of computing \code{e};
 
 \item
 \code{import "s"} imports a library of declarations from file \code{"s"}; subsequent import declarations for the \emph{same} file are not processed; furthermore, such imported files can only contain \code{let} or \code{import} commands.
\end{itemize}
}
\imgqool comes equipped with a set of built-in functions, among which arithmetic operators, logic primitives as described in Section \ref{sec:SpatialLogicFramework}, and imaging specific operators, stemming from basic tasks such as computing the gray-scale intensity of a colour image, or getting the separate colour components, to advanced operations such as computing cross correlation of histograms (see Section \ref{sec:DIASpecificOperators}). An exhaustive list of the available built-ins is provided in the user manual. Furthermore, a ``standard library'' is provided containing short-hands for commonly used functions, and for derived operators. 
An \textbf{expression} may be a numeric literal (no distinction is made between floating point and integer constants), a named constant (e.g. \code{x}), a function application (e.g. \code{f(x1,x2)}), an infix operator application (e.g. \code{x1 + x2}), or a parenthesized (sub-)expression (e.g. \code{(x1 + x2)}).

The language features \textbf{strong dynamic typing}, that is, types of expressions are unambiguously checked and errors are precisely reported, but such checks are only performed at ``run time'', that is, when evaluating closed-form expressions with no free variables.
No function or operator \emph{overloading} is possible in the current version, although this is a planned improvement to the type system, as well as some form of static typing. However, it is \emph{not} the case that a type error may waste a long-running analysis. Type checking occurs after loading and parsing, but before analysis is run. 

Actual program execution after parsing is divided into two phases. First (usually, in a negligible amount of time), all the ``save'' and ``print'' instructions are examined to determine what expressions actually \emph{need} to be computed; in this phase, name binding is resolved, all constant and function applications are substituted with closed expressions, types are checked and the environment associating names to expressions is discarded. Finally, the set of closed expressions to be evaluated is transformed into a set of tasks to be executed, possibly in parallel, and dependencies among them. After this phase, no further syntax processing or name resolution are needed, and it is guaranteed that the program is free from type errors. The second phase simply runs each task -- in dependency order --  parallelising execution on multiple CPU cores when possible.

Each built-in logical operator has an associated type of its input parameters and output result. The available types are inductively defined as \code{Number}, \code{Bool}, \code{String}, \code{Model}, and \code{Valuation(t)}, where \code{t} is in turn a type. The type \code{Model} is the type assigned to \code{x} in \code{load x = "f"}; operations such as the extraction of RGB components take this type as input, and return as output the only parametric type: \code{Valuation(t)}, which is the type of a multi-dimensional image in which each voxel contains a value of type \code{t}. For instance, the red component of a loaded model has type \code{Valuation(Number)}, whereas the result of evaluating a logic formula has type \code{Valuation(Bool)}\footnote{Although such type system would permit ``odd'' types such as \code{Valuation(Model)}, there is no way to construct them; in the future this may change when appropriate.}.

The design of \imgqool, and in particular its simple type system, and the low number of basic constructs, has been tailored to usability by an audience of users with diverse backgrounds. 
\ifArx 
For example, the tool abstracts from technical aspects such as the number of bits and representation (signed/unsigned integer or floating point) of numeric values, in favour of an extremely declarative and automated approach.
\fi 
In this respect, \imgqool expressions should be thought of as the equivalent, in the application domain, of SQL queries over a relational database. 
In this line, a very important aspect of the execution semantics of \imgqool specifications is the fact that expressions have no side effects and are amenable to \emph{memoization}, that is, intermediate results are cached and reused when computing equal sub-expressions. In \imgqool, memoization is not just an optimization 
technique, but rather the core of the execution engine (see Section \ref{sec:implementation-details}), used to achieve maximal sharing of subformulas along execution. In other words, in \imgqool, no expression is ever computed twice\footnote{Because of memoization, \imgqool might run out-of-memory; to avoid this, on-disk caching of results is typically used (also in the tool \topochecker, based on the very same theoretical foundations); although no experiments in our case studies have ever ran out of memory, on-disk caching is obviously planned in the short-term development road-map of \imgqool.}. This frees the user from having to worry about how many times a given function is called, and makes execution of complex macros and logical operators feasible. 
\ifArx 
To see this, consider that a function that uses one of its parameters twice, when called with an expression \code{e} as an argument, should in principle (e.g., when side-effects are allowed) evaluate \code{e} twice; this would lead to non-linear growth of the computation time with respect to the size of the input, which is avoided by memoization. Without using such technique, in our experiments, the number of subformulas to compute could easily reach numbers in the order of one million. With maximal sharing, this reduces to the order of one hundred. 
\fi 

\subsection{Efficiency and Comparison with \topochecker}

%
%
%
%
%
The evaluation of \imgqool in Sect.~\ref{sec:case-study} involves sets of 3D images of size $240\times240\times155$ (about 9 million voxels), and uses features of \imgqool, that are not present in \topochecker.  On the other hand, the example specification in~\cite{Ba+19}, and its variant aimed at 3D images, can be readily used to compare the performance of \imgqool and \topochecker. The specifications consist of two human-authored text files of about $30$ lines each; the largest logic formula in the experiments is the one identifying the \emph{oedema}, generating a syntax tree (when all macros are expanded) of depth about 80, and about one milion subformulas.  The machine used for testing is \computer. In the 2D case (image size: $512 \times 512$), \topochecker takes 52 seconds to complete the analysis, whereas \imgqool takes 750 milliseconds. In the 3D case (image size: $512\times512\times24$), \topochecker takes about 30 minutes, whereas \imgqool takes 15 seconds. As we mentioned already, such a huge improvement is due to the combination of a specialised imaging library, new algorithms (e.g., for statistical similarity of regions), parallel execution and other optimisations. One could get into more details by designing a specialised set of benchmarks, some of which can also be run using \topochecker; however, for the purposes of the current paper, the performance difference is so large that we do not deem such detailed comparison necessary. 

\subsection{Implementation details}\label{sec:implementation-details}

\imgqool is implemented in the functional, object-oriented programming language \code{FSharp}, using the \code{.NET Core} implementation of the \code{.NET} specification\footnote{See \url{https://fsharp.org} and \url{https://dotnet.github.io}}. This permits a single code base with minimal environment-dependent setup to be cross-compiled and deployed as a standalone executable, for the major desktop operating systems, namely \emph{Linux}, \emph{macOS}, and \emph{Windows}. 
%
Despite \code{.NET} code is compiled for an intermediate machine, this does not mean that efficiency of \imgqool is somehow ``non-native''. There are quite a number of measures in place to maximise efficiency. First and foremost, the execution time is heavily dominated by the time spent in native libraries (more details below), and \imgqool\ acts as a higher-level, declarative front-end for such libraries, adding a logical language, memoization, parallel execution, and abstraction from a plethora of technical details that a state-of-the-art imaging library necessarily exposes. In our experiments, parsing, memoization, and preparation of the tasks to be run may take a fraction of a second; the rest of the execution time (usually, several seconds, unless the analysis is extremely simple) is spent in \emph{foreign function calls}.

The major performance boosters in \imgqool are: a state-of-the-art computational imaging library (\code{ITK}); the optimised implementation of the \emph{may reach} operator; a new algorithm for statistical cross-correlation; an efficient memoizing execution engine; parallel evaluation of independent tasks, exploiting modern multi-core CPUs. Besides, special care has been put in making all performance-critical loops \emph{allocationless}. All used memory along the loops is pre-allocated, avoiding the risk to trigger garbage collection during computation. We will address each of them briefly in the following.

\paragraph{ITK library.} \imgqool is heavily based on the state-of-the-art imaging library \code{ITK}, exploiting the \code{SimpleITK} glue library\footnote{See \url{https://itk.org} and \url{http://www.simpleitk.org}}. Most of the logical and non-logical operators of \imgqool are implemented directly by a library call (notable exceptions include the \emph{reaches} logical connective, and statistical cross-correlation).

\paragraph{Novel algorithms.} The two most relevant operators that are not simply based on an \code{ITK} function are \code{mayReach} and \code{crossCorrelation}, implementing, respectively, the logical operator $\rho$, and statistical comparison described in Section~\ref{sec:DIASpecificOperators}. The computation of the voxels satisfying $\slreach{\phi_1}{\phi_2}$ can be implemented either using \emph{flooding} or by exploiting the connected components of $\phi_2$ as a flooding primitive; both solutions are available as primitives in \code{SimpleITK}, and in our experiments, connected components perform better using this library than plain flooding, for large input seeds. 
\ifArx 
More precisely, it is easy to see that the voxels satisfying $\slreach{\phi_1}{\phi_2}$ are those belonging either to $\lnear \phi_1$, or to $\lnear \psi$, where $\psi$ contains each voxel $x$ such that there is a connected component $C$ of $\phi_2$, with $x \in C$ and $C$ containing at least one voxel satisfying $\lnear \phi_1$.  
\fi 
The \emph{surrounded} logical connective is defined in terms of \code{mayReach}; also other frequently-used operators such as \code{touch} are derived from \emph{may reach}. Therefore, an optimised algorithm for \code{mayReach} is a key performance improvement. \code{CrossCorrelation} is a resource intensive operation, as it requires one to compute the histogram of a multi-dimensional hyperrectangle at each voxel. We note that it is not straightforward to apply pre-computation methods, such as the \emph{integral histogram}~\cite{Por05}, since the cross-correlation operator in \imgqool can be applied to images that are created along the computation by sub-formulas, so these are not known \emph{a priori}. In this work, we have designed a specialised parallel algorithm exploiting additivity of histograms. Given two sets of values $P_1$, $P_2$, let $h_1$, $h_2$ be their respective histograms, and let $h'_1$, $h'_2$ be the histograms of $P_1 \setminus P_2$ and $P_2 \setminus P_1$. For $i$ a bin, we have $h_2(i) = h_1(i) - h'_1(i) + h'_2(i)$. This property leads to a particularly efficient algorithm when $P_1$ and $P_2$ are two hyperrectangles centred over adjacent voxels, as $P_1 \setminus P_2$ and $P_2 \setminus P_1$ are \emph{hyperfaces}, having one dimension less than hyperrectangles. Thus, our algorithm first divides a multi-dimensional image into as many partitions as the number of processors in the machine running the program, and then computes a \emph{Hamiltonian path} $h^p$ for each partition $p$, passing by each voxel of \emph{p} exactly once (such paths are cached along program execution, too). After computing the histogram of a hypercube for each $h_p$ (namely the one centred on its first element), in parallel, each partition $p$ is visited in the order imposed by $h^p$, the histogram is computed incrementally as described above, and cross-correlation is also computed and stored in the resulting (quantitative) image, which can then be further manipulated, for example, by applying thresholds. 

\paragraph{Memoizing execution semantics.} As we already mentioned, no expression is computed twice in \imgqool. Sub-expressions are {\em by construction} identified up-to syntactic equality and assigned a number, representing a unique identifier (UID). UIDs start from $0$ and are contiguous, therefore admitting an array  of all existing sub-formulas to be used to pre-computed valuations of expressions without further hashing. 

\paragraph{Dereferentiation of memory pointers.} In performance-critical functions that do not have a direct counterpart in \code{ITK}, the interoperability-oriented functionality of \code{FSharp} (and the \code{dotnet} platform) is exploited to access the underlying memory of an image directly, even with array bounds checking turned off where possible, to achieve C-like efficiency in declarative, functional code. 

\subsection{Design and data structures}

The design of \imgqool defines three implementation layers. The \emph{core} execution engine implements the concurrent, memoizing semantics of the tool. The \emph{interpreter} is responsible for translating source code into core library invocations. These two layers 
only include some basic arithmetic and boolean primitives.
Operators can be added by inheriting from the abstract base class \code{Model}, which is part of the core layer, defining appropriate methods, and tagging them with the identifier and type to be used in the interpreter, using so-called \emph{\code{.NET} attributes}. Upon instantiation, all such methods are found via the \emph{reflection} capabilities of \code{FSharp}, and added to the ones available in the interpreter. 
The third implementation layer is the instantiation of the core layer to define operators from \SLCSMI, and loading and saving of graphical models, using the \code{ITK} library.  
We provide some more detail on the design of the core layer, which is the most critical part of \imgqool.
 At the time of writing, the core consists of just 350 lines of \code{FSharp} code, that has been carefully engineered not only for performance, but also for ease of maintenance and future extensions.   

A number of \textbf{classes} are essential to make this possible. The central ones are named \code{ModelChecker}, \code{FormulaFactory}, \code{Formula}, and \code{Operator}, of which \code{Constant} is a subclass. Class \code{Operator} is used for evaluating an operator with a provided list of arguments (via the method \code{Eval}). Its attributes include a string name, and the type of arguments and return value.  The class \code{Formula} is a symbolic representation of a syntactic sub-expression. Each instance of \code{Formula} has a unique numeric id (UID), an instance of \code{Operator}, and (inductively) a list of \code{Formula} instances, denoting its arguments. The UID of a formula is determined by the operator name (which is unique across the application), and the list of parameters. Therefore, by construction, it is not possible to build two different instances of \code{Formula} that are syntactically equal. The class \code{FormulaFactory} manages UID assignment. It features a \code{Create} method, that given an instance of \code{Operator} and a list of UIDs, returns either a fresh instance of \code{Formula} or an existing one, using a hash table for the purpose. 
\ifArx In the current implementation, there is only one instance of \code{FormulaFactory}, but nothing prevents one to use more instances in the same program for different instances of the core layer. 
\fi
UIDs are contiguous and start from $0$. This permits all created formulas to be inserted into an array. Furthermore, UIDs are allocated in such a way that the natural number order is a topological sort of the dependency graph between subformulas (that is, if $f_1$ is a parameter of $f_2$, the UID of $f_1$ is greater than the UID of $f_2$). This is exploited in class \code{ModelChecker}; internally, the class uses an array to store the results of evaluating each \code{Formula} instance, implementing memoization. The class \code{ModelChecker} turns each formula in a task to be executed. Whenever formula with UID $i$ is an argument of the formula with UID $j$, a dependency is noted between the associated tasks. The class \code{ModelChecker} then makes use of the high-level, lightweight concurrent programming library \code{Hopac}\footnote{See \url{https://github.com/Hopac/Hopac}.} and its abstractions to maximise CPU usage on multi-core machines, by queuing tasks for parallel evaluation whenever their dependencies have already been evaluated. \code{Hopac} takes care of scheduling tasks to processor cores. This guarantees high CPU efficiency\footnote{For instance, on the machine with 4 cores that we employ for our tests,  we obtain about $350\%$ CPU usage or even more, depending on the experiment (the less the dependencies, the more the efficiency). It is well known that such kind of task scheduling is in general a hard problem, which may be tackled more specifically in future work; however, execution times in current applications are quite satisfactory and do not make such optimisations a high priority task.}. The interpreter uses one instance of the class \code{ModelChecker}, which is parameterised by one instance of \code{Model}. 
\ifArx
In \imgqool, only one instance of \code{Model} is defined, namely the ``third layer'' specialised to image processing that we already mentioned.
\fi

\ifArx
\section{Experimental Evaluation:  Segmentation of Glioblastoma in 3D Medical Images}
\else
\section{Experimental Evaluation}
\fi
\label{sec:case-study}

%

We have evaluated the performance of  \imgqool\ in two ways. The first considers the use of \imgqool\ for the segmentation of Glioblastoma in 3D medical images obtained as MRI scans in NIfTI-1 format from patients in preparation for radiotherapy. In radiotherapy the {\em clinical target volume} (CTV) of the whole tumour is considered, which is an extension of the {\em gross tumour volume} (GTV), corresponding to what can actually be seen on an image. For glioblastomas this margin is a 2-2.5 cm isotropic expansion of the GTV volume within the brain. 

\ifArx 
In addition, the performance of \imgqool\ has been evaluated on the Brain Tumor Image Segmentation Benchmark (BraTS) of 2017~\cite{Menze2015,Bak+17} for what concerns the GTV of the whole tumour. This evaluation provides information on the quality of the proposed approach compared to the most advanced techniques that are currently being developed.
The 2017 version of this benchmark contains 211 multi contrast MRI scans of low and high grade glioma patients that have been obtained from multiple institutions and were acquired with different clinical protocols and various scanners. All the imaging data sets provided by \brats have been segmented manually and were approved by experienced neuro-radiologists. In our evaluation we have used the T2 Fluid Attenuated Inversion Recovery (FLAIR) type of scans, which is one of the four provided modalities in the benchmark. 
\else
In addition, the performance of \imgqool\ has been evaluated on 211 multi contrast MR scans\footnote{We have used the T2 Fluid Attenuated Inversion Recovery (FLAIR) type of scans, which is one of the four modalities provided in the benchmark. 17 of them have been excluded because they contained some form of clearly identifiable artefact in the FLAIR image, that currently is the only type of image we consider.} of the Brain Tumor Image Segmentation Benchmark (BraTS) of 2017~\cite{Menze2015,Bak+17} for what concerns the GTV of the whole tumour. 
\ifArx
This evaluation provides information on the quality of the proposed approach compared to the most advanced techniques that are currently being developed. 
\fi 
\fi

\subsection{\ImgQL\ segmentation procedure}

\begin{figure}[!t]
\centering
\subfloat[][] 
{
\includegraphics[height=2.0cm]{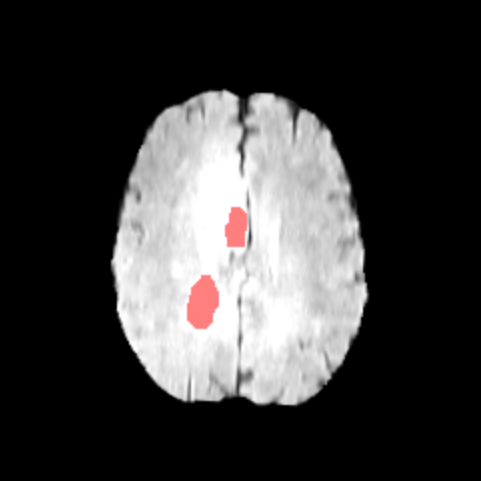}
\label{subfig:hyper}
}\hfill
\centering
\subfloat[][] 
{
\includegraphics[height=2.0cm]{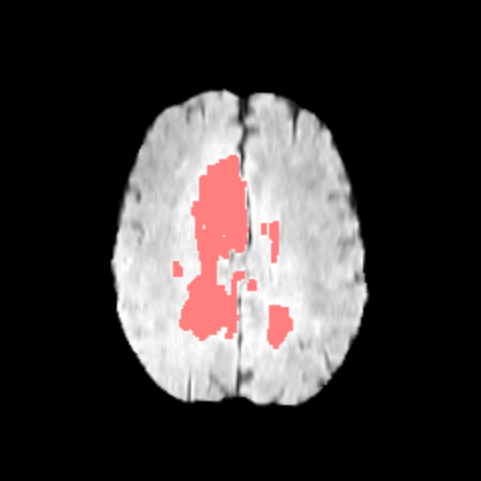}
\label{subfig:very}
}\hfill
\centering
\subfloat[][] 
{
\includegraphics[height=2.0cm]{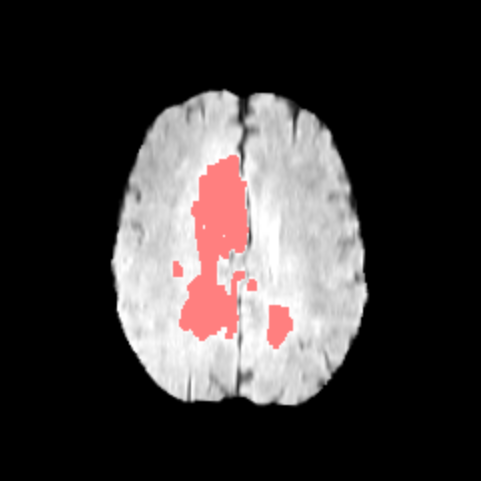}
\label{subfig:gTum}
}\hfill
\centering
\subfloat[][] 
{
\includegraphics[height=2.0cm]{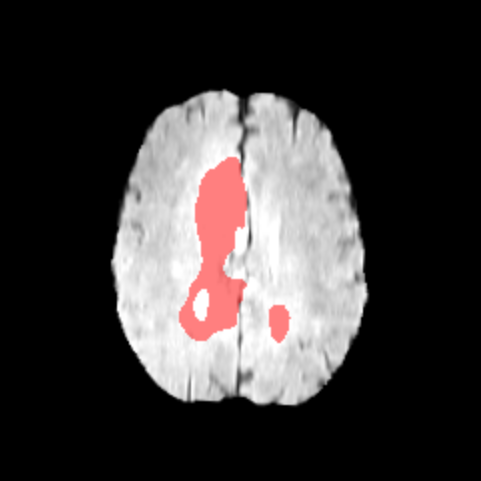}
\label{subfig:stat}
}\hfill
\centering
\subfloat[][]
{
\includegraphics[height=2.0cm]{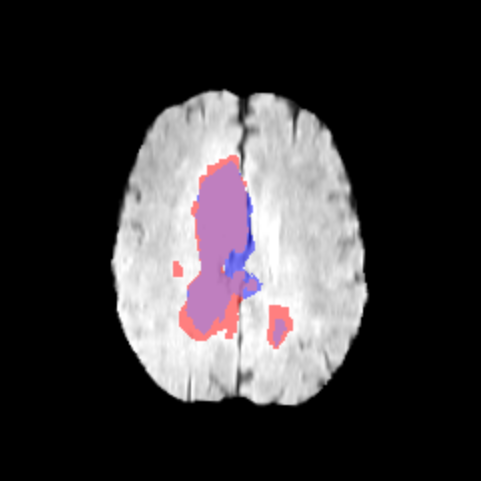}
\label{subfig:complete}
}
\caption{Tumour segmentation of image {\tt Brats17\_2013\_2\_1}, FLAIR, axial 2D slice at X=155, Y=117 and Z=97. (a) hyperIntense (b) veryIntense (c) growTum (d) tumStatCC (e)  tumFinal (red) and manual (blue, overlapping area is purple).}
\label{fig:seg}
\end{figure}

Specification~\ref{alg:seg} shows the tumour segmentation procedure that we used for the evaluation. The syntax is that of \imgqool, namely: \code{|,\&,!} are boolean \emph{or}, \emph{and}, \emph{not}; \code{distlt(r,a)} is the set of points having distance less than \code{r} from the points that are true in \code{a} (similarly, \code{distgt}; distances are in millimiters); \code{crossCorrelation(r,a,b,phi,m,M,k)} is an intermediate function yielding a cross-correlation coefficient for each voxel, to which a predicate $c$ may be applied to obtain the \emph{statistical similarity} function of Section \ref{sec:DIASpecificOperators}; the \code{>} operator performs thresholding of an image; \code{border} is true on voxels that lay at the border of the image; other operators should be self-explaining.   The various intermediate phases of the segmentation procedure, for axial view of one specific 2D slice of an example 3D MRI scan of the BraTS 2017 data set, are shown in Fig~\ref{fig:seg}.
\ifArx 
We briefly discuss the main parts of this specification. Line 1 imports the definitions of some basic derived operators. Lines 2-4 define a few specific operators for the segmentation (see Def.~\ref{def:derived}). Lines 4-5 load the 3D MRI scan in .nii format as well as the manually segmented one used for calculating the indexes. Line 9 defines the background i.e. all voxels having intensity less than 0.1 which are part of an area that touches the border of the image. Points that are part of the brain are all those points that do not satisfy \code{background}.
The application of  \code{percentiles} in line 11 assigns to each point of the brain the percentile rank to which it belongs considering the range of intensities of points that are part of the brain in the FLAIR image. Based on these percentiles hyper-intense and very-intense points are identified that satisfy hI and vI, respectively (Lines 12-13). Hyper-intense points have a very high likelihood to belong to tumour tissue and the very-high intensity points are likely to belong to the tumour as well, or to the oedema that is usually surrounding the tumour. However, not all hyper-intense and very-intense points are part of a tumour. The idea is to identify the real tumour using further information. First of all, the hyper-intense points should form an area of certain dimensions and moreover they should be close to the area covered by the oedema. 

Such characteristics can be easily specified in the \ImgQL\ language. In lines 14-15 the hyper-intense and very-intense points are filtered such that noise is removed (i.e. single points or very small areas of hyper-intense points) and only areas of a certain relevant size are considered. The points that satisfy \code{hyperIntense} and \code{veryIntense} are shown in red in Fig.~\ref{subfig:hyper} and in Fig~\ref{subfig:very}, respectively. In line 16 the areas of hyper-intense points are extended via the \code{grow} operator with those points that are very intense and part of a contiguous area of very intense points (the oedema that accompanies the tumour) that in turn touches the hyper-intense areas (likely tumour tissue). The points that satisfy \code{growTum} are shown in red in Fig.~\ref{subfig:gTum}.
In line 17 the previously-defined (line 8) similarity operator is used to assign to all voxels a texture-similarity score with respect to \code{growTum}. In line 18 this operator is used to find those voxels that have a high cross correlation coefficient and likely part of the tumour. The result is shown in Fig.~\ref{subfig:stat}. Finally (line 19), the voxels that are identified as part of the whole tumour are those that satisfy \code{growTum} extended with those that are statistically similar to it via the \code{grow} operator. Points that satisfy \code{tumFinal} are shown in red in Fig.~\ref{subfig:complete} and points identified by manual segmentation are shown for comparison in blue in the same figure (so that the overlapping areas are shown in purple).
\fi 
\begin{algorithm}
    \tt
\SetAlgoLined
\caption{\label{alg:seg}\ImgQL\ specification of tumour segmentation}
\imgImport "stdlib.imgql"\\[1em]
\imgLet grow(a,b) = (a \imgOr touch(b,a))\\ 
\imgLet flt(r,a) = distlt(r,distgeq(r,\imgNot a))\\
\imgLoad imgFLAIR =  "Brats17\_2013\_2\_1\_flair.nii.gz"\\ 
\imgLoad imgManualSeg = "Brats17\_2013\_2\_1\_seg.nii.gz"\\ 
\imgLet manualContouring = intensity(imgManualSeg) $>$ 0\\[1em]
\imgLet flair = intensity(imgFLAIR)\\
\imgLet similarFLAIRTo(a) = crossCorrelation(5,flair,flair,a,min(flair),max(flair),100)\\[1em]
\imgLet background = touch(flair $<$ 0.1,border)\\
\imgLet brain = \imgNot background\\
\imgLet pflair = percentiles(flair,brain)\\[1em] 
\imgLet hI = pflair $>$ 0.95\\
\imgLet vI = pflair $>$ 0.86\\
\imgLet hyperIntense = flt(5.0,hI)\\
\imgLet veryIntense =  flt(2.0,vI)\\[1em]
\imgLet growTum = grow(hyperIntense,veryIntense)\\
\imgLet tumSim = similarFLAIRTo(growTum)\\
\imgLet tumStatCC = flt(2.0,(tumSim $>$ 0.6))\\
\imgLet tumFinal= grow(growTum,tumStatCC)\\[1em]
\ifArx 
// Compute indexes\\
\imgLet truePositives = volume(tumFinal \imgAnd manualContouring)\\
\imgLet trueNegatives = volume((\imgNot tumFinal) \imgAnd (\imgNot manualContouring))\\
\imgLet falseNegatives = volume((\imgNot tumFinal) \imgAnd manualContouring)\\
\imgLet falsePositives = volume(tumFinal \imgAnd (\imgNot manualContouring))\\
\imgLet sensitivity = truePositives / (truePositives + falseNegatives)\\
\imgLet specificity = trueNegatives / (trueNegatives + falsePositives)\\
\imgLet dice = (2 * truePositives) / ((2 * truePositives) + falseNegatives + falsePositives)\\[1em]
\fi
// Save results\\
\imgSave "output\_Brats17\_2013\_2\_1/complete-FLAIR\_FL-seg.nii" tumFinal
\end{algorithm}

Interesting aspects of the \ImgQL\ specification are its relative simplicity and abstraction level, fitting that of neuro-radiologists, its explainability, its time-efficient verification, admitting a rapid development cycle, and its independence of normalisation procedures through the use of percentiles rather than absolute values for the intensity of voxels.

\subsection{\imgqool\ evaluation results}
%
%

In this domain results of tumour segmentation are evaluated based on a number of indexes commonly used to compare the quality of different techniques. These indexes are based on the true positive (TP) voxels (voxels that are identified as part of a tumour in both manual and \imgqool\ segmentation), true negatives (TN) voxels (those that are {\em not} identified as part of a tumour in both manual and \imgqool\ segmentation), false positives (FP) voxels (those identified as part of a tumour by \imgqool\ but not by manual segmentation) and false negatives (FN) voxels (those identified as part of a tumour by manual segmentation but not by \imgqool). Based on these four types the following indexes are defined: 
\ifArx 
\begin{itemize}
\item sensitivity: $\mbox{TP} / (\mbox{TP} + \mbox{FN})$
\item specificity: $\mbox{TN} / (\mbox{TN} + \mbox{FP})$
\item Dice: $2*\mbox{TP} / (2*\mbox{TP} + \mbox{FN} +\mbox{FP})$
\end{itemize}
\else 
{\em sensitivity}: $\mbox{TP} / (\mbox{TP} + \mbox{FN})$; {\em specificity}: $\mbox{TN} / (\mbox{TN} + \mbox{FP})$; {\em Dice}: $2*\mbox{TP} / (2*\mbox{TP} + \mbox{FN} +\mbox{FP})$.
\fi
%
%
\ifArx 
Sensitivity measures the fraction of voxels that are correctly identified as part of a tumour. Specificity measures the fraction of voxels that are correctly identified as {\em not} being part of a tumour.
The Dice similarity coefficient is used to provide a measure of the similarity of two segmentations and ranges from 0.0 to 1.0, where 0.0 indicates that there is no overlap and 1.0 indicates a perfect overlap between the manual and \imgqool\ segmentation results. 
\fi 
%
%
Table~\ref{tab:results} shows the mean values of the above indexes for both GTV and CTV volumes for Specification~\ref{alg:seg} applied to the BraTS 2017 training phase collection. For comparison, we report also the mean and standard deviation of the 20 techniques reported in the BraTS 2017 Segmentation Challenge~\cite{brats17preproc} that were applied to at least 100 cases of the same training phase data. Full data for these techniques is available in the leaderboard\footnote{https://www.cbica.upenn.edu/BraTS17/lboardTraining.html}.
%
%
\begin{table}
\begin{center}
\begin{tabular}{| l | r | r | r | r | r | r | } \hline
\multicolumn{4}{|c|}{\imgqool\ Mean (stdev) of 194 cases} & \multicolumn{3}{c|}{BraTS17 Mean (stdev) of 20 techniques}\\ \hline
         & Sens. & Spec. & Dice &  Sens. & Spec. & Dice \\ \hline 
GTV & 0,884(0,107)& 0,998(0,002)& 0,849(0,098)  &  0.846(0.120) & 0.992(0.005) & 0.850(0.095) \\ \hline
CTV & 0,948(0,069)& 0,991(0,013)& 0,905(0,085) & N.A.& N.A.& N.A.\\ \hline
\end{tabular}
\end{center}
\caption{\label{tab:results}\imgqool\ evaluation results on the BraTS 2017 benchmark. }
\end{table}
%
%
%
%
Table~\ref{tab:results} shows that our results are well in line with the state-of-the-art in this domain, which is very encouraging. The evaluation of each case study takes about 10 seconds on \computer, paving the way to interactive use and visualisation by neuro-radiologists, which is one of the longer term aims of our work.
For the CTV a Dice value of 0.9 is sufficiently accurate for use in preparation of radiotherapy. 
\ifArx 
Further inspection of results of individual cases may provide ideas for further improvements of Specification~\ref{alg:seg}.  
Furthermore, in this evaluation we have focused on the segmentation of the tumour as a whole, leaving further segmentation of the tumour in various other types of tissue (e.g. necrotic and non-enhancing parts) for future research.
\fi 


%% file: Conclusions.tex
\section{Conclusions and Future Work}\label{sec:Conclusions}

We presented \imgqool, a spatial model checker designed and optimised for the analysis of multi-dimensional digital images. The tool has been successfully evaluated on 211 cases of an international brain tumour 3D MRI segmentation benchmark. The obtained results are well-positioned w.r.t. the performance of state-of-the-art segmentation techniques, both efficiency-wise and accuracy-wise. 
\ifArx
Concerning efficiency, the \ImgQL\ specification that has been evaluated provides results on 3D MRI images in less than 15 seconds on a desktop machine. The results are well inline with the current state-of-art for accuracy based on commonly accepted indexes, which is exceptional for the simple specification provided (less than 30 lines including loading, saving and printing indexes), that is open for further improvement.
\fi
Future research work based on the tool will focus on further benchmarking (e.g. various other types of tumours and tumour tissue such as necrotic and non-enhancing parts), and clinical application. On the development side, planned future work includes a graphical (web) interface for interactve parameter calibration (for that, execution times will need to be further improved, possibly employing \emph{GPU computing}); improvements in the type-system (e.g. \emph{operator overloading}); turning the first two design layers into a reusable library available for other projects. Finally, the (currently small, albeit useful) library of logical and imaging-related primitives available will be enhanced, based on input from case studies. Experimentation in combining \emph{machine-learning} technologies with the logic-based approach of \imgqool are also a worthwile research line to explore.

%% file: TACAS19.bbl
\begin{thebibliography}{10}
\providecommand{\url}[1]{\texttt{#1}}
\providecommand{\urlprefix}{URL }

\bibitem{HBSL}
Aiello, M., Pratt{-}Hartmann, I., van Benthem, J.: Handbook of Spatial Logics.
  Springer (2007)

\bibitem{Bak+17}
Bakas, S., Akbari, H., Sotiras, A., Bilello, M., Rozycki, M., Kirby, J.S.,
  Freymann, J.B., Farahani, K., Davatzikos, C.: Advancing the cancer genome
  atlas glioma mri collections with expert segmentation labels and radiomic
  features. Scientific Data  4 (2017),
  \url{https://doi.org/10.1038/sdata.2017.117}, online publication date:
  2017/09/05

\bibitem{Ba+19}
{Banci Buonamici}, F., Belmonte, G., Ciancia, V., Latella, D., Massink, M.:
  {Spatial Logics and Model Checking for Medical Imaging} (Submitted for
  publication; available also as \RED{XARKIVE})

\bibitem{BBLN17}
Bartocci, E., Bortolussi, L., Loreti, M., Nenzi, L.: Monitoring mobile and
  spatially distributed cyber-physical systems. In: Proceedings of the 15th
  ACM-IEEE International Conference on Formal Methods and Models for System
  Design. pp. 146--155. MEMOCODE '17, ACM, New York, NY, USA (2017),
  \url{http://doi.acm.org/10.1145/3127041.3127050}

\bibitem{Bartocci2016}
Bartocci, E., Gol, E.A., Haghighi, I., Belta, C.: A formal methods approach to
  pattern recognition and synthesis in reaction diffusion networks. {IEEE}
  Transactions on Control of Network Systems pp. 1--1 (2016),
  \url{https://doi.org/10.1109%2Ftcns.2016.2609138}

\bibitem{Be+17}
Belmonte, G., Ciancia, V., Latella, D., Massink, M., Biondi, M., Otto, G.D.,
  Nardone, V., Rubino, G., Vanzi, E., Buonamici, F.B.: A topological method for
  automatic segmentation of glioblastoma in mr flair for radiotherapy -
  {ESMRMB} 2017, 34th annual scientific meeting. Magnetic Resonance Materials
  in Physics, Biology and Medicine  30(S1),  437 (oct 2017),
  \url{https://doi.org/10.1007/s10334-017-0634-z}

\bibitem{BCLM16}
Belmonte, G., Ciancia, V., Latella, D., Massink, M.: From collective adaptive
  systems to human centric computation and back: Spatial model checking for
  medical imaging. In: ter Beek, M.H., Loreti, M. (eds.) Proceedings of the
  Workshop on FORmal methods for the quantitative Evaluation of Collective
  Adaptive SysTems, FORECAST@STAF 2016, Vienna, Austria, 8 July 2016. {EPTCS},
  vol. 217, pp. 81--92 (2016), \url{http://dx.doi.org/10.4204/EPTCS.217.10}

\bibitem{Ch5HBSL}
van Benthem, J., Bezhanishvili, G.: Modal logics of space. In: Handbook of
  Spatial Logics, pp. 217--298. Springer (2007)

\bibitem{Castellano2004}
Castellano, G., Bonilha, L., Li, L., Cendes, F.: Texture analysis of medical
  images. Clinical Radiology  59(12),  1061--1069 (dec 2004)

\bibitem{CGLLM14}
Ciancia, V., Gilmore, S., Latella, D., Loreti, M., Massink, M.: Data
  verification for collective adaptive systems: Spatial model-checking of
  vehicle location data. In: Eighth {IEEE} International Conference on
  Self-Adaptive and Self-Organizing Systems Workshops, {SASOW}. pp. 32--37.
  {IEEE} Computer Society (2014)

\bibitem{CGLLM15}
Ciancia, V., Grilletti, G., Latella, D., Loreti, M., Massink, M.: An
  experimental spatio-temporal model checker. In: Software Engineering and
  Formal Methods - {SEFM} 2015 Collocated Workshops. Lecture Notes in Computer
  Science, vol. 9509, pp. 297--311. Springer (2015)

\bibitem{CLLM14}
Ciancia, V., Latella, D., Loreti, M., Massink, M.: Specifying and verifying
  properties of space. In: Theoretical Computer Science - 8th {IFIP} {TC} 1/WG
  2.2 International Conference, {TCS} 2014, Rome, Italy, September 1-3, 2014.
  Proceedings. Lecture Notes in Computer Science, vol. 8705, pp. 222--235.
  Springer (2014)

\bibitem{CLLM16}
Ciancia, V., Latella, D., Loreti, M., Massink, M.: {Model Checking Spatial
  Logics for Closure Spaces}. {Logical Methods in Computer Science}  {Volume
  12, Issue 4} (Oct 2016), \url{http://lmcs.episciences.org/2067}

\bibitem{CLMP15}
Ciancia, V., Latella, D., Massink, M., Pakauskas, R.: Exploring spatio-temporal
  properties of bike-sharing systems. In: 2015 {IEEE} International Conference
  on Self-Adaptive and Self-Organizing Systems Workshops, {SASO} Workshops. pp.
  74--79. {IEEE} Computer Society (2015)

\bibitem{Ciancia2018}
Ciancia, V., Gilmore, S., Grilletti, G., Latella, D., Loreti, M., Massink, M.:
  Spatio-temporal model checking of vehicular movement in public transport
  systems. International Journal on Software Tools for Technology Transfer
  (Jan 2018), \url{https://doi.org/10.1007/s10009-018-0483-8}

\bibitem{CLLM16bertinoro}
Ciancia, V., Latella, D., Loreti, M., Massink, M.: Spatial logic and spatial
  model checking for closure spaces. In: Bernardo, M., Nicola, R.D., Hillston,
  J. (eds.) Formal Methods for the Quantitative Evaluation of Collective
  Adaptive Systems - 16th International School on Formal Methods for the Design
  of Computer, Communication, and Software Systems, {SFM} 2016, Bertinoro,
  Italy, June 20-24, 2016, Advanced Lectures. Lecture Notes in Computer
  Science, vol. 9700, pp. 156--201. Springer (2016)

\bibitem{CLMPV16}
Ciancia, V., Latella, D., Massink, M., Paskauskas, R., Vandin, A.: A tool-chain
  for statistical spatio-temporal model checking of bike sharing systems. In:
  Margaria, T., Steffen, B. (eds.) Leveraging Applications of Formal Methods,
  Verification and Validation: Foundational Techniques - 7th International
  Symposium, ISoLA 2016, Imperial, Corfu, Greece, October 10-14, 2016,
  Proceedings, Part {I}. Lecture Notes in Computer Science, vol. 9952, pp.
  657--673 (2016), \url{http://dx.doi.org/10.1007/978-3-319-47166-2_46}

\bibitem{Davnall2012}
Davnall, F., Yip, C.S.P., Ljungqvist, G., Selmi, M., Ng, F., Sanghera, B.,
  Ganeshan, B., Miles, K.A., Cook, G.J., Goh, V.: Assessment of tumor
  heterogeneity: an emerging imaging tool for clinical practice? Insights into
  Imaging  3(6),  573--589 (oct 2012)

\bibitem{Despotovi2015}
Despotovi{\'{c}}, I., Goossens, B., Philips, W.: {MRI} segmentation of the
  human brain: Challenges, methods, and applications. Computational and
  Mathematical Methods in Medicine  2015,  1--23 (2015),
  \url{http://dx.doi.org/10.1155/2015/450341}

\bibitem{Dupont2016}
Dupont, C., Betrouni, N., Reyns, N., Vermandel, M.: On image segmentation
  methods applied to glioblastoma: State of art and new trends. {IRBM}  37(3),
  131--143 (jun 2016), \url{https://doi.org/10.1016%2Fj.irbm.2015.12.004}

\bibitem{brats17preproc}
(Ed.), S.: 2017 international miccai brats challenge: pre-conference
  proceedings (09 2017),
  \url{https://www.cbica.upenn.edu/sbia/Spyridon.Bakas/MICCAI_BraTS/MICCAI_BraTS_2017_proceedings_shortPapers.pdf}

\bibitem{Fyllingen2016}
Fyllingen, E.H., Stensj{\o}en, A.L., Berntsen, E.M., Solheim, O., Reinertsen,
  I.: Glioblastoma segmentation: Comparison of three different software
  packages. {PLOS} {ONE}  11(10),  e0164891 (oct 2016),
  \url{https://doi.org/10.1371%2Fjournal.pone.0164891}

\bibitem{Gal99}
Galton, A.: The mereotopology of discrete space. In: Freksa, C., Mark, D.
  (eds.) Spatial Information Theory. Cognitive and Computational Foundations of
  Geographic Information Science, Lecture Notes in Computer Science, vol. 1661,
  pp. 251--266. Springer Berlin Heidelberg (1999),
  \url{http://dx.doi.org/10.1007/3-540-48384-5_17}

\bibitem{Gal03}
Galton, A.: A generalized topological view of motion in discrete space. Theor.
  Comput. Sci.  305(1-3),  111--134 (2003),
  \url{https://doi.org/10.1016/S0304-3975(02)00701-6}

\bibitem{Gal14}
Galton, A.: Discrete mereotopology. In: Calosi, C., Graziani, P. (eds.)
  Mereology and the Sciences: Parts and Wholes in the Contemporary Scientific
  Context, pp. 293--321. Springer International Publishing (2014),
  \url{https://doi.org/10.1007/978-3-319-05356-1_11}

\bibitem{Gr+09}
Grosu, R., Smolka, S., Corradini, F., Wasilewska, A., Entcheva, E., Bartocci,
  E.: Learning and detecting emergent behavior in networks of cardiac myocytes.
  Commun. ACM  52(3),  97--105 (2009)

\bibitem{SPATEL}
Haghighi, I., Jones, A., Kong, Z., Bartocci, E., Grosu, R., Belta, C.: Spatel:
  A novel spatial-temporal logic and its applications to networked systems. In:
  Proceedings of the 18th International Conference on Hybrid Systems:
  Computation and Control. pp. 189--198. HSCC '15, ACM, New York, NY, USA
  (2015)

\bibitem{Kassner2010}
Kassner, A., Thornhill, R.E.: Texture analysis: A review of neurologic {MR}
  imaging applications. Am. J. Neuroradiol.  31(5),  809--816 (2010)

\bibitem{lemieux1999fast}
Lemieux, L., Hagemann, G., Krakow, K., Woermann, F.: Fast, accurate, and
  reproducible automatic segmentation of the brain in t1-weighted volume mri
  data. Magnetic Resonance in Medicine  42(1),  127--135 (1999)

\bibitem{Lopes2011}
Lopes, R., Ayache, A., Makni, N., Puech, P., Villers, A., Mordon, S., Betrouni,
  N.: Prostate cancer characterization on {MR} images using fractal features.
  Med. Phys.  38(1), ~83 (2011)

\bibitem{Menze2015}
Menze, B.H.e.a.: The multimodal brain tumor image segmentation benchmark
  (brats). IEEE Transactions on Medical Imaging  34(10),  1993--2024 (2015)

\bibitem{NBCLM15}
Nenzi, L., Bortolussi, L., Ciancia, V., Loreti, M., Massink, M.: Qualitative
  and quantitative monitoring of spatio-temporal properties. In: Runtime
  Verification - 6th International Conference, {RV} 2015 Vienna, Austria,
  September 22-25, 2015. Proceedings. Lecture Notes in Computer Science, vol.
  9333, pp. 21--37. Springer (2015)

\bibitem{Por05}
Porikli, F.M.: Integral histogram: a fast way to extract histograms in
  cartesian spaces. 2005 IEEE Computer Society Conference on Computer Vision
  and Pattern Recognition (CVPR'05)  1,  829--836 vol. 1 (2005)

\bibitem{Simi2015}
Simi, V., Joseph, J.: Segmentation of glioblastoma multiforme from {MR} images
  {\textendash} a comprehensive review. The Egyptian Journal of Radiology and
  Nuclear Medicine  46(4),  1105--1110 (dec 2015),
  \url{https://doi.org/10.1016%2Fj.ejrnm.2015.08.001}

\bibitem{TKG17}
Tsigkanos, C., Kehrer, T., Ghezzi, C.: Modeling and verification of evolving
  cyber-physical spaces. In: Proceedings of the 2017 11th Joint Meeting on
  Foundations of Software Engineering. pp. 38--48. ESEC/FSE 2017, ACM, New
  York, NY, USA (2017), \url{http://doi.acm.org/10.1145/3106237.3106299}

\bibitem{Zhu2012}
Zhu, Y., Young, G.S., Xue, Z., Huang, R.Y., You, H., Setayesh, K., Hatabu, H.,
  Cao, F., Wong, S.T.: Semi-automatic segmentation software for quantitative
  clinical brain glioblastoma evaluation. Academic Radiology  19(8),  977--985
  (aug 2012), \url{https://doi.org/10.1016%2Fj.acra.2012.03.026}

\end{thebibliography}
